\documentclass[12pt]{amsart} 
\usepackage{moreverb}

\usepackage[colorlinks,bookmarksopen,bookmarksnumbered,citecolor=red,urlcolor=red]{hyperref}

\usepackage[T1]{fontenc}
\usepackage[utf8]{inputenc}

\usepackage{amsmath}
\usepackage{amsfonts}
\usepackage{amsthm}

\usepackage{dsfont}
\usepackage{hhline}
\usepackage{booktabs}
\usepackage{graphicx,xcolor}
\usepackage{subfig}
\usepackage{rotating}
\usepackage{multirow}
\usepackage{epstopdf}

\definecolor{vert1}{rgb}{0.,0.50,0.2}

\DeclareMathOperator*{\argmin}{\arg\min}

\newcommand\BibTeX{{\rmfamily B\kern-.05em \textsc{i\kern-.025em b}\kern-.08em
T\kern-.1667em\lower.7ex\hbox{E}\kern-.125emX}}

\begin{document}

%\runauthor{A. Fischer et al.}
%
%\runtitle{Statistical learning for wind power}

%\articletype{RESEARCH ARTICLE}

\title[Statistical learning for wind power]{Statistical learning for wind power : a modeling and stability study towards forecasting}
\author[A. Fischer et al.]{Aurélie Fischer, Lucie Montuelle, Mathilde Mougeot \& Dominique Picard}
\address{Laboratoire de Probabilités et Modèles Aléatoires, Université Paris Diderot, 75013 Paris, France}

%\corraddr{L. Montuelle, Laboratoire de Mathématiques Appliquées, Agrocampus Ouest, 65 rue de Saint-Brieuc, CS 84215, 35042 Rennes Cedex, France.\\
%Email: lucie.montuelle@agrocampus-ouest.fr}

\footnote{Funding : Agence Nationale de la Recherche (ANR-14-CE05-0028)}

%\cgs{Agence Nationale de la Recherche (ANR-14-CE05-0028)}

\begin{abstract}
%short-term wind power forecast
		We focus on wind power modeling using machine learning techniques. We show on real data provided by the wind energy company Ma\"ia Eolis, that parametric models, even following closely the physical equation relating wind production to wind speed are outperformed by intelligent learning algorithms. 
		In particular, the CART-Bagging algorithm gives very stable and promising results.
		Besides, as a step towards forecast, we quantify the impact of using deteriorated wind measures  on the  performances.
		We show also on this application that the default methodology to select a subset of predictors provided in the standard  random forest package  can be refined,  especially when there exists among the predictors one variable which has a major impact.
		\end{abstract}
		

\maketitle
		
\section{Introduction}

In France,  wind energy represents today 3.9\% of the national electricity production.
The United Nations Conference on Climate Change COP21 has set a goal of 30\%  renewable energy in the overall energy supply in the country by 2020, 
and more precisely,  the French wind production should double by 2020 \cite{Panorama2015}.

Since electricity can hardly be stored,  forecasting tools are essential  to appropriately balance  the production of  the different renewable energies.
Today, in France, wind energy is produced by more than 1400 wind farms  scattered all over the country.
The production of each wind farm is highly dependent of the meteorological conditions and especially of the wind. It is well known that the behavior of the wind is very different from one region to another, and this seems especially significant in France, where several quite different climates are present despite the relatively small area of the country \cite{Panorama2015}. 
So, to be accurate, the global wind electricity forecast should rely on local models, dedicated to each wind farm. 
Consequently, an important  first step is to quantify the modeling performances of wind production in the different French regions,  using real operational data.

Two kinds of framework are usually investigated  today  for wind power prediction. On the one side, physical models rely on  the modeling
of each wind turbine based on  equations \cite{costa2008review}. 
On another side, a trend of new mathematical tools  tends to model  the power production by  learning the phenomenon directly on the data, without integrating any knowledge on the physical behavior of the wind turbines. Such techniques using  
statistical models and data mining methods have been investigated in many complex situations, for instance considering short term prediction. Among others, parametric regression models, Support Vector Machines  for regression, regression trees, random forests, neural networks have been considered. 
For instance,  the use of neural networks has been investigated in \cite{sideratos2012probabilistic, kramer2013wind} and in \cite{quan2014short}. % with predictions intervals. 
A special network, called  extreme learning machine, has been used in \cite{wan2014probabilistic} for probabilistic interval forecasting.  %to provide probabilistic forecast.
%Prediction intervals are also provided in \cite{wan2014probabilistic}, thanks to extreme learning machine probabilistic forecasting. 
The $k$-nearest neighbor method has been studied for wind power modeling by \cite{kusiak2009wind}.  
In \cite{Mangalova2016}, the $k$-nearest neighbor algorithm is used for probabilistic forecasts in the frame of the 
Global Energy Forecasting Competition 2014.

%Probabilistic forecast is also provided in \cite{wan2014probabilistic}, thanks to extreme learning machine.  
  Support vector machines for regression have been proposed in this context in  \cite{kramer2011short}, whereas \cite{kusiak2009wind} provides a comparison between several data-mining approaches.
Besides, time series-based models have also contributed to the field of wind power forecast (see, e.g., \cite{michael2004statistical,wu2014wind}). For an overview of different modeling and forecasting methods for wind power, the reader may further refer to the surveys  \cite{costa2008review,foley2012current,giebel2011state, jung2014current}.
	%---see also the references therein}.

%Recently, probabilistic forecasts for wind power generation have been introduced during the 
%Global Energy Forecasting Competition 2014 \cite{Mangalova2016}.

In the present paper, adopting the second point of view, we investigate and compare different techniques for modeling  the electrical power for several wind farms in France.
For each farm, we first model  the electrical power of each wind turbine of the farm
 using local inputs  coming from sensors directly installed on each wind turbine.
The predictive power of the farm is then given by the sum of the predictive powers computed for each wind turbine.
%and then forecasting
In a second step, we quantify the modeling performances by using more global inputs
as may be provided by a meteorologist forecaster as for example, Météo France. 
This approach helps to quantify   the performance of  the different  models 
running in an operational environment, using only average input information at a farm scale.
%real time forecast

The CART-Bagging algorithm appears to perform the best on our data and gives very satisfactory predictions. 

The paper is organized as follows. In Section \ref{section:data}, we thoroughly describe the data set at hand. Section \ref{section:meth} introduces the different methods investigated in our study. Section \ref{section:resloc} presents and discusses the modeling performances obtained using the local information on each turbine. The results found when replacing this information by the more global  one, relying on averages, are given in Section \ref{section:resglob}.

\section{Data set} \label{section:data}

		The data set has been provided by the wind energy company Maïa Eolis.
In a farm, each wind turbine provides 10 minute measurements of electrical power, wind speed, wind direction, temperature, as well as an indicator of the working state of the turbine. The electrical power output of the whole farm is also provided on a 10 minute basis.  All measures are recorded simultaneously. Data is  available for 3 different farms made up of 4 to 6 turbines, in the North and East of France, from 2011 to 2014.

To detect freeze, wind speed is measured  on each turbine both by a classical anemometer and a heated one. Since more measures are available from the heated anemometer, the study has been conducted with this data. 
Wind direction is provided by a weather vane and has been recoded into two variables corresponding to the cosine and the sine of the angle.
The state of the turbine may correspond to start, stop or full working of the turbine,  depending on the wind speed and maintenance operations. For the sake of simplicity, this study focuses on fully operating times. Besides, the data has been averaged over 30 minutes in order to slightly smooth the signals. However, it should be stressed that most often the results obtained  on a 10 minute basis are quite similar to those presented in the sequel.

		Taking advantage of the 30 minutes averages, two new variables have been introduced: the variance of the wind speed, and the variance of the wind direction over 30 minutes. The second variable (complex-valued), has been decomposed into its real part and its imaginary part, leading to a total of 7 explanatory variables.

\section{Predictive methods} \label{section:meth}

			In this section, all the measures are assumed to be observed in real time. Based on this data, our aim is to \emph{model} the farm power. More precisely, the variables are observed at time $t$ and the sum of the power of each turbine of the farm at time $t$ is predicted. We recall that the studied model is applied at each turbine, providing an estimate of its power. Then the estimated farm power is computed by summing the estimated turbine powers. The error is calculated at the farm scale.
		
		Our intention is to compare parametric statistical methods closely reflecting the related physical equation, to more elaborated techniques inspired from machine learning. These methods are especially designed to learn a phenomenon in a completely agnostic way and may be  suitable for high dimensional data or complex data. In particular, they can easily accommodate non-linear modeling as well as  dependence between variables, which is the case here.
%forecast

\subsection{Theoretical equation}

\begin{figure}
\centering
\subfloat{\label{fig:PowerCurve-a} \includegraphics[width=.5\textwidth]{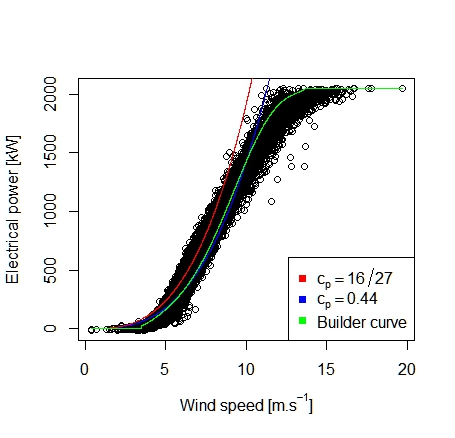} }
\subfloat{\label{fig:PowerCurve-cube} \includegraphics[width=.5\textwidth]{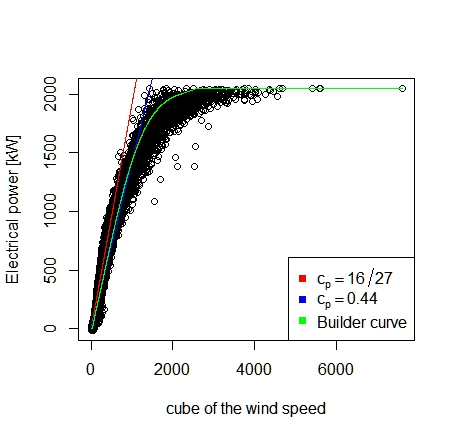} }
\caption{Empirical observations for a wind turbine and theoretical power curves for different power coefficient values, compared to the curve provided by the turbine builder. }
\label{fig:PowerCurve}
\end{figure}

		According to theoretical studies on wind turbines (see, e.g., \cite{Lydia2014452}), the delivered power obeys the following equation : 
		\begin{equation}
		 P(W) = \frac{1}{2} \rho S c_p W^3 , 
		 \label{eq:power}
		\end{equation}
where $W$ is the wind speed, $\rho$ the air density, $S$ the rotor surface, which is the area swept by the blades, and $c_p$ the power coefficient, corresponding to the fraction of wind energy that the wind turbine is able to extract. Thus, as expected, the power significantly depends on the wind speed and a good approximation of the power curve could lead to good predictions using wind speed measurements. Figure~\ref{fig:PowerCurve} shows the raw observations and the fit to the theoretical curves for a wind turbine. Figure~\ref{fig:PowerCurve-a} plots power versus wind-speed, whereas Figure~\ref{fig:PowerCurve-cube} plots power versus the cube of the wind speed.  The two plotted theoretical curves  correspond to two different values of $c_p$: the maximal theoretical value ($16/27$, red curve), and a more realistic value given in Table~8 of \cite{Carrillo2013572} (blue curve). The third curve (in green) is provided by the turbine builder, based on his experiments.

 Notice that the cloud of observations is quite dispersed and we can already anticipate difficulties for prediction.
 
In particular, it should be underlined that the parameters of the physical equation~\eqref{eq:power} are in practice difficult to guess, so that the theoretical curves may not fit very well. Furthermore, to better reflect the observations, the theoretical formula is often used only for a range of wind speeds, outside which the power is assumed to be constant. However, the knowledge of this range requires the estimation of both endpoints of the interval.

Although these curves  correspond to some trend, there is obviously room for improvement to produce a better prediction.

\subsection{Parametric methods}

Several methods have been tested to approximate the power curve and model the production.
In this section, we present the parametric statistical methods, directly inspired from the physical equation. 
%forecast

\paragraph{Parametric modeling according to  the wind speed only}

We first investigated the  simplest parametric  models, namely linear regression and logistic regression, with the wind speed as unique explanatory variable. 
If the predicted power at time $t$ is denoted by $\hat{Y}_t$, the linear model is given by
\[ \hat{Y}_t=  \hat a_0+\hat a_1 W_t, \] 
where $W_t$ denotes the wind speed at time $t$, and $\hat a_0$ and $\hat a_1$ are computed using ordinary least squares (OLS).
The logistic regression model may be written
\[ \hat{Y}_t= \frac{\hat C}{1+\exp (\hat a_0+\hat a_1 W_t)}, \]
where the parameters $\hat a_0, $ $\hat a_1$, $\hat C$  are obtained using maximum likelihood.
% \com{Lucie: à préciser}

 Introducing a third degree polynomial of the wind speed in the logistic regression has also been considered to mimic more closely Equation~\eqref{eq:power}. More precisely, the model is then defined by:  
\[ \hat{Y}_t= \frac{\hat C}{1+\exp (\hat a_0+\hat a_1 W_t + \hat a_2 W_t^2 + \hat a_3 W_t^3)}, \] where $\hat a_i $, $i=0,\dots, 3$ and $\hat C$ are estimated parameters.
This model is denominated  in the sequel as  polynomial logistic regression.

\subsubsection{Parametric modeling using more variables}

Linear regression, logistic regression and polynomial logistic regression have also been  studied with more variables, using not only wind speed as a predictor, but also wind direction, (coded by its cosine and sine : $D^{\cos}$ and  $D^{\sin}$), temperature $T$ and the variances of the wind speed $W^S$ and direction, $D^{S,\mathfrak{Re}}$ and $D^{S,\mathfrak{Im}}$. Denoting for each variable $X$ by $X_t$ its value at time $t$, the corresponding equations describing the estimated power $\hat Y_t$ may be written as follows.

\begin{align}
\begin{split} \label{eq:lin2}
\hat{Y}_t &=  \hat a_0+ \hat a_1 W_t+ \hat a_2 D^{\cos}_t+ \hat a_3 D^{\sin}_t+ \hat a_4 T_t+ \hat a_5 W_t^S + \hat a_6 D^{S,\mathfrak{Re}}_t \\ &+ \hat a_7 D^{S,\mathfrak{Im}}_t 
\end{split} \\
\begin{split}
\hat{Y}_t &= \hat C \Big(1+\exp( \hat a_0+ \hat a_1 W_t+ \hat a_2 D^{\cos}_t+ \hat a_3 D^{\sin}_t+ \hat a_4 T_t+ \hat a_5 W_t^S \\ 
&+ \hat a_6 D^{S,\mathfrak{Re}}_t+ \hat a_7 D^{S,\mathfrak{Im}}_t )\Big)^{-1} 
\end{split}
\\
\begin{split}
\hat{Y}_t &= 
\hat C\left(1+\exp\left( \hat a_0+ \sum_{k=1}^{3}\hat a_{1,k} (W_t)^k+ \hat a_2 D^{\cos}_t+ \hat a_3 D^{\sin}_t+ \hat a_4 T_t \right. \right. \\& \left. \left.
+ \hat a_5 W_t^S + \hat a_6 D^{S,\mathfrak{Re}}_t+ \hat a_7 D^{S,\mathfrak{Im}}_t \right)\right)^{-1} 
\end{split}
\end{align}
In the last equation, corresponding to polynomial logistic regression, only the wind speed $W_t$ occurs in the expression as a polynomial of order 3, to be consistent with the theoretical equation \eqref{eq:power}. All the parameters of the different methods are estimated as in the previous paragraph.

 The Lasso method, which simultaneously performs variable selection and regularization through the least squares criterion penalized by the $\ell^1$ norm of the regression coefficients has been investigated as well (see for instance \cite{Lasso}). For this model, the predicted power at time $t$ is a linear combination of all the previous variables as in equation \eqref{eq:lin2}, the coefficients $\hat a_1,\dots, \hat a_7$ being estimated with OLS, under the usual constraint $\sum_{j=1}^7 |a_j|\leq \kappa$ for some constant $\kappa >0$.
 
 %Here, the regularization parameter  $\lambda$ is calibrated, as usual, by cross-validation on the data.
 
 %\[ \hat{Y}_t=  \hat a_0+ \hat a_1 W_t+ \hat a_2 D^{\cos}_t+ \hat a_3 D^{\sin}_t+ \hat a_4 T_t+ \hat a_5 W_t^S + \hat a_6 D^{S,\mathfrak{Re}}_t+ \hat a_7 D^{S,\mathfrak{Im}}_t ,\]with $\hat a_0, \ldots, \hat a_7$
 
% minimizing \[\frac{1}{n} \sum_{i=1}^n \left(Y_i-a_0-a_1 W_i- a_2 D^{\cos}_i- a_3 D^{\sin}_i- a_4 T_i- a_5 W_i^S - a_6 D^{S,\mathfrak{Re}}_i- a_7 D^{S,\mathfrak{Im}}_i \right)^2+ \lambda \sum_{j=1}^7 |a_j|.\]

\subsection{Non-parametric methods and machine learning algorithms}

It is well-known that non-parametric and non-linear methods are very useful to model complex phenomena.  The following algorithms do not generally lead to closed formulas as in the previous section. We will describe them briefly and refer to the literature for more details.

\subsubsection{KNN}
The $k$-nearest-neighbor procedure consists in computing the average power corresponding to the $k$ nearest neighbors in the feature space  (see for instance \cite{80546}). More precisely, the rule is given by
$$\hat Y_t=\frac 1k\sum_{j=1}^k Y_{(j)},$$ where $Y_{(j)}$ corresponds to the wind power of the $j$-th nearest neighbor of the  observation at time $t$, according to the Euclidean distance of the variables at time $t$. The features are standardize to have mean 0 and variance 1 since they are measured in different units. 

As for the previous method, the  number $k$ of neighbors is  optimized on a grid.

\subsubsection{CART, Bagging and RF}  Tree-based methods like CART~\cite{MR726392} and Random Forests~\cite{RFBreiman} are also applied. CART grows a binary tree by choosing at each step the cut minimizing the intra-node variance, over all explanatory variables already introduced in equation \eqref{eq:lin2} (here denoted by $X_j$, $j=1,\dots, 7$) and all possible  thresholds (denoted by $S_j$ hereafter). More specifically, the intra-node variance, usually called deviance, is defined by $$ D(X_j,S_j)=\sum_{X_j<S_j}(Y_s-\overline{ Y}^-)^2+\sum_{X_j\geq S_j}(Y_s-\overline{ Y}^+)^2,$$ where $\overline{ Y}^-$ (respectively $\overline{ Y}^+$) denotes the average of the wind power in the  area $\{X_j<S_j\}$ (respectively $\{X_j\geq S_j\}$).
Then, the selected $j_0$ variable and associated threshold is given by $(X_{j_0}, S_{j_0})=\argmin_{j,S_j}D(X_j,S_j)$.
 To avoid over-fitting, the tree is usually pruned by cross-validation.
The prediction is provided by the  value associated to  the leaf in which the observation falls.

Another way to reduce variance and avoid over-fitting is to use Bagging~\cite{Breiman96baggingpredictors}. Bagging consists in generating bootstrap samples, fitting a method on every sample (here growing a full tree by CART) and averaging the predictions.
Hence, for $B$ bootstrap samples, the predicted power is given by \begin{equation}\label{eq:bag}
\hat Y_t=\sum_{b=1}^{B}\hat Y_t^b,
\end{equation}
where $\hat Y_t^b$ denotes the power predicted by CART for the $b$-th bootstrap sample.
To produce more diversity in the trees to be averaged, an additional random step may be introduced in the previous procedure, leading to Random Forests. 
In the Random Forests procedure, each tree is grown following the same principle as in CART (with no pruning), but, here, the best cut is chosen among a smaller subset of randomly chosen variables. The predicted value is the mean of the predictions of the trees, as in \eqref{eq:bag}.

\subsubsection{SVM for regression} The SVM method for regression  maps the inputs into a non-linear feature space, using a kernel representation, for example a Gaussian kernel (see for instance \cite{NIPS1996SVM}). A non-linear regression function is computed  by  minimizing the sum of the losses on the points giving rise to an error exceeding some threshold. The threshold parameter is here calibrated   using a grid.

\bigskip

In the next section, all the experiments have been conducted using the R software. The previous procedures are implemented respectively in the packages \verb?lars?,  \verb?kernlab?,  \verb?FNN?, \verb?rpart? and \verb?randomForest?. 
For the Random Forests, the default parameters, advocated by Breiman, were used: 500 trees were grown in each forest and the size of the subset of randomly chosen variables, commonly denoted by \emph{mtry}, is the floor of the third of the number of variables. Note that the CART-Bagging  algorithm is a particular case of Random Forests where \emph{mtry} equals the total number of variables.

\subsection{The naive method}

Finally, the so-called  ``persistence method'' uses the last observation as prediction: if $Y_t$ denotes the electric production at time $t$,  the predicted production at time $t$ is given by $\hat{Y}_t = Y_{t-1} $. It is interesting to introduce this very naive method as a benchmark in comparison to more sophisticated methods to precisely quantify their gain.

\subsection{From turbine to farm modeling}

As mentioned, the evaluation of the performances is made at the farm scale. Therefore, each turbine is modeled using the evaluated method, then the estimation of each wind turbine power is provided on test points. Finally, the estimated power of the farm is computed by summing theses estimations. More precisely, if the farm comprises six turbines and the linear regression is considered, six linear regression models are adjusted,  then predictions for the test set are computed on each turbine: $\hat{Y}_{t,1}, \hat{Y}_{t,2},\hat{Y}_{t,3},\hat{Y}_{t,4},\hat{Y}_{t,5},\hat{Y}_{t,6}$, and finally the estimation of the farm power is given by $\hat{Y}_t=\sum_{i=1}^6 \hat{Y}_{t,i}$.

%\section{Real Time Forecasting results}\label{section:resloc}

\section{Modeling performance results}\label{section:resloc}
As we are interested in evaluating the predictive power of each method, the data set is split as usual into a training and a test set. 
In order to quantify the variability of the predictive ability, several test  sets are used. An average performance, as well as a standard deviation, are then computed.
 	
		More precisely, the procedures are trained on around $8 000$ instant-points and 10 data sets of 724 points are used to evaluate the performances. The error criterion is the Root Mean Squared Error (RMSE), defined between a vector of predictions $\hat{Y}$ and a vector of observed wind power productions $Y$ by 
		\[ RMSE(\hat{Y}) = \sqrt{ \frac{1}{T} \sum_{t=1}^T (\hat{Y}_t - Y_t)^2} .
		\]
A quantity which is also of interest for industries is the error in term of percentage of the installed power (\% of IP in the results tables), defined by the average RMSE divided by the theoretical power of the farm. For example, if the farm is composed of 6 turbines of theoretical power $2.05$ GW (specified by the turbines builder), the error in term of percentage of the installed power is \[\% \mbox{ of IP }= 100 \times \frac{mean(RMSE)}{6\times 2050} \%.\] This quantity sometimes appears under the denomination Normalized RMSE.

\begin{table}
\caption{Modeling performances using local measures for one farm (IP= Installed Power).}
%\caption{Performances for real time forecast procedures with local measures for one farm (IP= Installed Power).}
\label{table:instant_methods}
\centering
\small
\begin{tabular}{ l c c c c }
%\toprule
  & Method & Mean of RMSE & Sd of RMSE & \% of IP \\
\midrule
 & Persistence & 855.52 & 141.14 & 6.96 \\
\hline
\multirow{6}{*}{ \begin{sideways}using wind\end{sideways} \begin{sideways} speed only \end{sideways} }  & 
 Linear Regression & 373.61 & 86.91 & 3.04 \\ 
& Logistic Regression & 404.86 & 76.74 & 3.29 \\ 
& Polynomial Log. Reg. & 290.36 & 73.87 & 2.36 \\
& CART & 314.46 & 57.74 & 2.56 \\ 
& \textbf{CART-Bagging (=RF)} & \textbf{250.52} & \textbf{46.52} & \textbf{2.04} \\ 
& SVM for regression & 269.94 & 64.21 & 2.19 \\ 
\hline
\multirow{9}{*}{ \begin{sideways} using all variables \end{sideways} }  & Linear Regression & 364.21 & 102.39 & 2.96 \\ 
& Logistic Regression & 362.76 & 107.58 & 2.95 \\ 
& Polynomial Log. Reg. & 292.57 & 100.53 & 2.38 \\
& LASSO& 364.21 & 102.39 & 2.96  \\
& CART & 314.46 & 57.74 & 2.56 \\
& \textbf{CART-Bagging}  & \textbf{203.50} & \textbf{39.72} & \textbf{1.65} \\ 
& RF & 425.78 & 161.53 & 3.46 \\ 
& SVM for regression & 382.16 & 134.34 & 3.11 \\ 
& kNN (k=2) & 355.29 & 109.96 & 2.89 \\
%\bottomrule
\end{tabular}
\end{table}

\begin{figure}
\centering
\includegraphics[width=\textwidth]{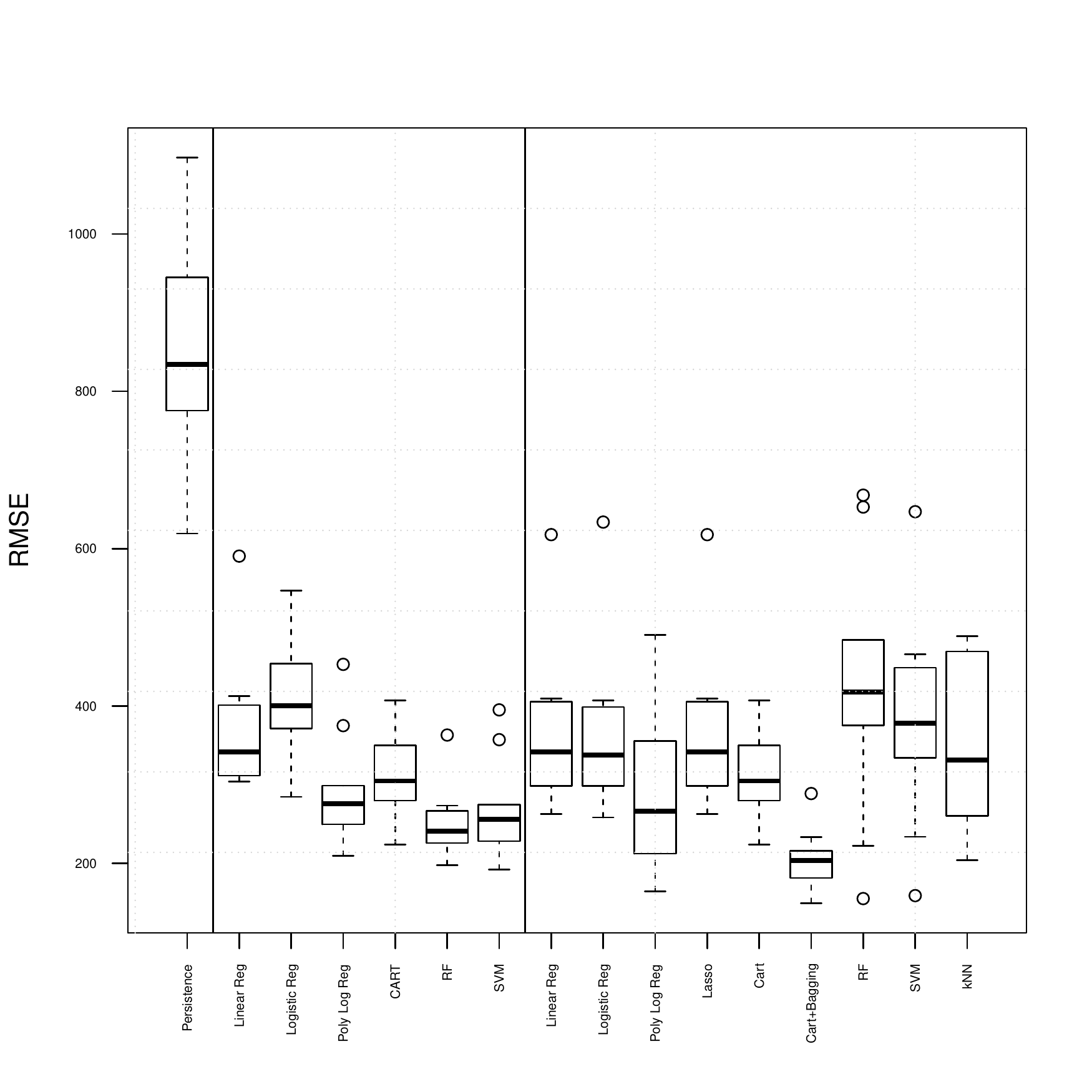}
\caption{Boxplots of the RMSE for the different procedures using local measures for one farm.}
\label{fig:Boxplot_Vobs}
\end{figure}		

Let us comment the main conclusions drawn thanks to Table \ref{table:instant_methods} and Figure \ref{fig:Boxplot_Vobs}.

\subsubsection{General observations}

As can be observed in Table \ref{table:instant_methods}, the learning algorithms have been investigated either using the wind speed variable only, in which case the emphasis is on the non-linear added value of the method,  or using all variables, insisting then on both the non-linear and regularization aspects. 
\\
We first observe  that all the methods investigated show a much better performance than the naive persistence method, substantially reducing the mean error,  with a standard deviation almost always better.

\subsubsection{Wind speed only}

 Concerning the methods using only the wind speed  as predictor, their performances are pretty good,  more than twice better than persistence. 

 The  polynomial logistic regression shows a very good performance, which was expected since this model is directly inspired from the physical equations as illustrated in Figure~\ref{fig:PowerCurve}. However, the variability of the prediction is a bit high.
 
The SVM and RF methods for regression show the best results with the best stability. 
Note that in that particular case, RF and the CART-Bagging procedure coincide.

\subsubsection{All variables}
Regarding the parametric methods, the results show that adding more variables, namely the wind direction, the variances of the wind speed and direction, and the temperature,  do not lead to any substantial  improvement. Among these procedures, polynomial logistic regression shows the best performances.

The LASSO procedure is not very promising. This is probably due to multiple factors :  the method uses the predictors in a linear way -- compared to SVM or CART, which are bringing different kinds of non-linearity -- and  the predictors are  highly  correlated. We observe also that the results are the same for  LASSO and the classical linear regression due to the fact that no selection has been in fact performed by the method : all the variables are kept.

It appears that the CART algorithm does not take advantage of the additional variables and seem to choose its cuts only according to the wind speed. This may be explained by the prevailing importance of the wind speed over other measures. 

Among the agnostic machine learning algorithms, the SVM shows one of the poorest performances.  It should be noted that several tested kernels were not able to compete with the polynomial logistic regression for example.

The kNN method has a performance similar to the SVM procedure.

The CART-Bagging algorithm  outperforms all the investigated statistical models.
	The case of Random Forests is quite interesting and has to be discussed separately. Looking at the Table~\ref{table:instant_methods} and Figure~\ref{fig:Boxplot_Vobs}, we can observe that RF surprisingly seem  less efficient than other methods and especially CART when dealing with all variables. However this poor result has to be refined. 

As explained above, the RF algorithm, instead of considering all the variables to grow a tree (as CART does), operates a random selection among these variables. The default choice for this random selection is the uniform distribution to choose a subset of the original variables, of size \emph{mtry}, the floor of the third of the number of predictors. In the  CART-Bagging procedure, all the variables are selected. In our data set, obviously, the importance of the wind speed prevails over all other variables: for instance, CART performs nearly all its cuts according to the wind speed. Therefore, if the wind speed variable is often not selected in a random sample, the resulting cut is often not appropriate. Choosing more variables increases the probability to select a specific variable, namely, here, the wind speed. 
Very different performances are then observed between RF with the default parameter for \emph{mtry} and the CART-Bagging method, corresponding to RF with \emph{mtry} equal to the number of predictors.

Comparing CART and CART-Bagging highlights the advantages of bootstrapping and averaging. This step allows to reduce the error by a third, when dealing with all the predictors.

 Note that, according to the renewable energy union~\cite{Panorama2015}, French industries obtain a root mean squared error of 2.4~\% of the installed power  of farm productions, which illustrates the benefits of using CART-Bagging (1.65~\%).
%for real-time forecast

\subsubsection{Comparison of different farms }

The results given in the previous paragraphs concern a farm in the East of France. Data from two different sites in the North of France were also available. For every farm, the hierarchy between procedures is quite similar, the procedure ranking first  most often is CART-Bagging.

To make a fair comparison between the farms, a new experiment has been conducted. A common test set, with observed variables available at the same time for each farm, with at least one turbine fully operational, has been drawn. The test set has been divided into ten subsets of 1440 instant-points, each covering a period of around thirty days, to quantify the average performance and its variability.
The training set consists in around 7200 instant-points, satisfying a ratio of $83 \%$ of the data dedicated to learning and $17 \%$ used for test.

\begin{figure}
\centering
\includegraphics[width=0.5\textwidth]{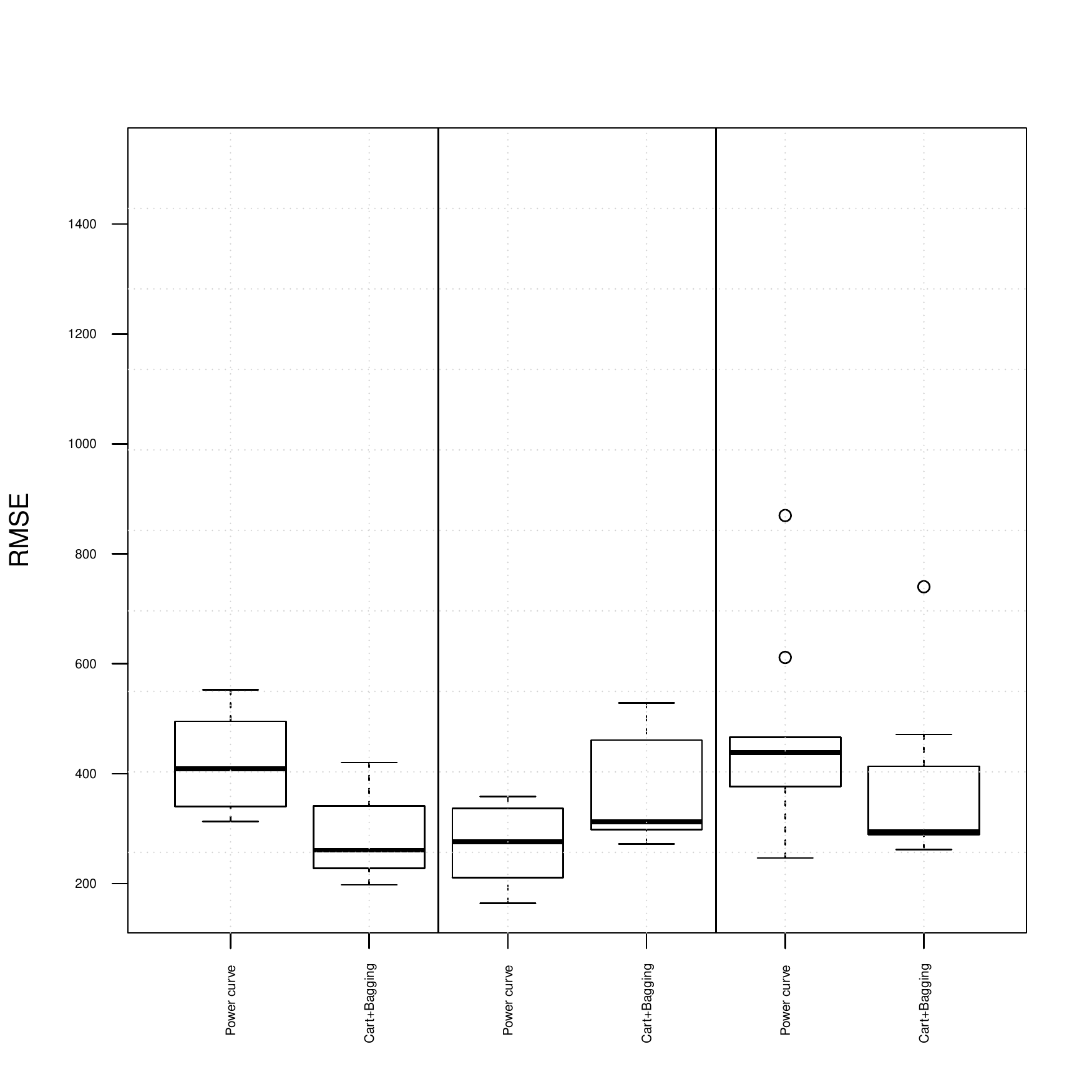}
\caption{Comparison of the RMSE for the turbine builder's power curve and the CART-Bagging procedure on several farms using local measures.}
\label{fig:ComparaisonParcsVobs}
\end{figure}

Only the best procedure, CART-Bagging, has been applied. We also compare the results with  the turbine builder's power curve, used on each turbine to model the farm and represented by the green curve in Figure~\ref{fig:PowerCurve}. Figure~\ref{fig:ComparaisonParcsVobs} highlights the good results of CART-Bagging on the first and the third farms. It performs reasonably well on the second farm, but is not as good as  the power curve's builder. It may be explained by the difference between the wind speed in the training sample and in the test set. Few high wind speed levels are observed in the training sample on the second farm compared to the test sample, so the CART-Bagging prediction may not be accurate.

%\section{Forecast framework} \label{section:resglob}
\section{Towards forecast : a stability investigation} \label{section:resglob}

%Our article focuses mainly on comparing statistical and machine learning methods for modeling the link between the wind (and other meteorological variables) and the electrical power generated in a wind farm. Indeed, 
%As explained in the paper, such data were not available in the frame of this work.  
%That is why we decided to introduce deteriorated virtual sensors (averages of true wind information) to study the robustness of the different procedures to  the use of data that is less precise and less spatially localized, just as meteorological forecasts. Interestingly, we observe clearly that the nonparametric method CART-Bagging outperforms all other methods regarding this stability study. Therefore, we argue that CART-Bagging should be a good choice for providing electrical forecasts in combination with appropriate meteorological wind forecast inputs.
%
%%%%%%%%%%%%%%%%%%

Forecasting electrical power requires two steps: one is to provide forecasts of the explanatory variables, and the other one, which is our aim here, consists in constructing an accurate model to plug the previous forecasts in. Indeed, if we have at hand an efficient  model, then the performance of a 
forecasting procedure of electrical wind power will directly depend on  wind forecasts. Up to now, the best wind forecasts are directly inspired from physics: these meteorological forecasts are obtained thanks to ensemble methods based on numerical computations, mostly based on the Navier-Stokes equations, like the climate reanalyzes performed for instance by  the European Centre for Medium-Range Weather Forecasts (ECMWF), or, for France, the French Weather Agency Météo France. 
%%%%%%%

At a farm scale, on a daily use, many observations are recorded in real time on each wind turbine. For example, as already mentioned, each turbine has its own anemometer and vane, which provide very localized information about wind speed and wind direction. Thanks to the analysis conducted in the previous section, we have been able to identify an accurate model, built with this kind of observations. It should be noted that, in a wind farm, in general, two wind turbines are at a distance of about 300~m from each other. However, concerning numerical models, for instance, the finest grid resolution for forecast of wind and temperature provided by the French Weather Agency Météo France is brought by the AROME model, which proposes a  resolution  of about $1.5$ km (5 times larger).
 Consequently, an interesting question is also to quantify the predictive power not using  very local information, but  information on a much broader scale.

To mimic the scale of meteorological wind forecasts, which are in the frame of this project not available, we decided to introduce virtual sensors: for each variable, a global information is computed by averaging all the localized variables coming from the set of turbines installed on the wind farm. Studying, thanks to this kind of data, which is less precise and less spatially localized than the true wind information, the stability of the different procedures,  helps to quantify the loss of accuracy due to the replacement of all the localized data with a unique global information and is a first step towards forecast.

\begin{figure}
\centering
\includegraphics[width=\textwidth]{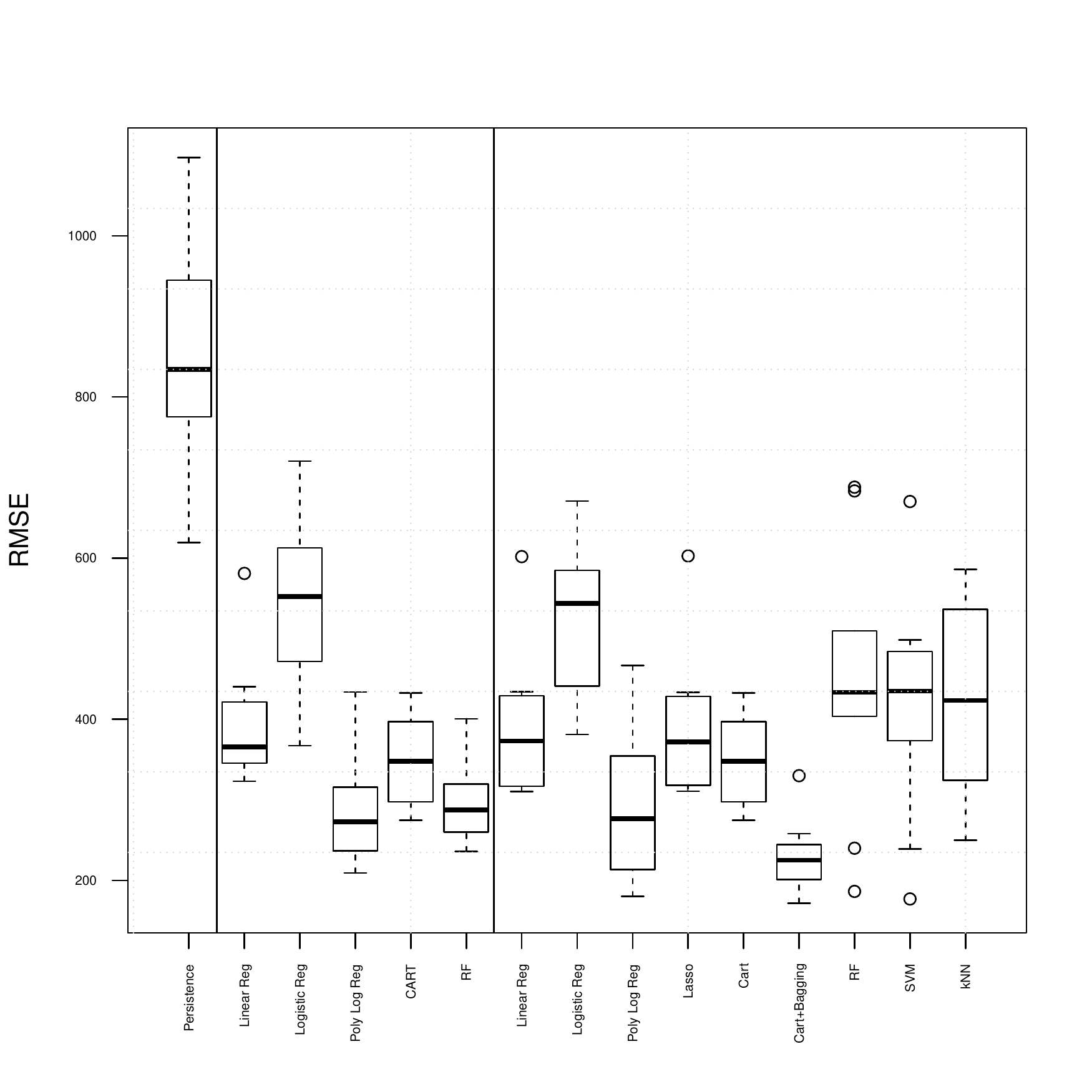}
\caption{Boxplots of the RMSE for the different procedures using the mean of  wind speed measures.}
\label{fig:Boxplot_Vmed}
\end{figure}

	The same methods have been used and the results are available in Table~\ref{table:instant_methods_Vmed}. The deterioration of the prediction can easily be seen in Figure~\ref{fig:Boxplot_Vmed}. We observe that polynomial logistic regression is remarkably robust, performing similarly to the context with local measures, contrary to SVM and kNN. When only wind speed is considered, polynomial logistic regression competes with  CART-Bagging, whereas the latter outperforms all the considered procedures when dealing with all the variables.
		Therefore, we argue that CART-Bagging should be a good choice for providing electrical forecasts in combination with appropriate meteorological wind forecast inputs.
	\\

%	As can be seen in Table \ref{table:instant_methods_Vmed}, the behavior of a polynomial logistic regression with only the wind speed as covariate is at the same time simple and effective for power prediction. 
%	It is interesting to notice that agnostic methods are in this case very appropriate  for prediction and show promising  results with the best stability, particularly for CART-Bagging.

\begin{table}
\caption{Modeling performances using deteriorated wind measures (average).}
%\caption{Performances for real time forecast procedures with the mean of wind measures.}
\label{table:instant_methods_Vmed}
\centering
\small
\begin{tabular}{ l c c c c}
\toprule
  & Method & Mean of RMSE & Sd of RMSE & \% of IP \\
\midrule
 & Persistence & 855.52 & 141.14 & 6.96 \\
\hline
\multirow{5}{*}{ \begin{sideways}using wind\end{sideways} \begin{sideways} speed only \end{sideways} }  & 
 Linear Regression &  393.09 & 77.25 & 3.20 \\ 
& Logistic Regression & 541.37 & 103.15 & 4.40 \\ 
& \textbf{Polynomial Log. Reg.} & \textbf{288.28} & \textbf{75.23} & \textbf{2.34} \\
& CART & 349.17 & 53.20 & 2.84 \\
& CART+Bagging (=RF) & 293.26 & 48.96 & 2.38 \\ 
\hline
\multirow{9}{*}{ \begin{sideways} using all variables \end{sideways} }  & Linear Regression &  387.71 & 89.73 & 3.15 \\ 
& Logistic Regression & 524.30 & 92.58 & 4.26 \\ 
& Polynomial Log. Reg. &  297.16 & 92.79 & 2.42 \\
& LASSO & 387.44 & 89.86 & 3.15 \\
& CART & 349.17 & 53.20 & 2.84  \\
& \textbf{CART + Bagging} & \textbf{228.75} & \textbf{43.35} & \textbf{1.86} \\ 
& RF & 447.77 & 161.84 & 3.64 \\ 
& SVM & 424.15 & 143.02 & 3.45 \\ 
& kNN & 428.05 & 125.84 & 3.48 \\
\bottomrule
\end{tabular}
\end{table}

\subsubsection{Comparison of different farms}
\begin{figure}
\centering
\includegraphics[width=.5\textwidth]{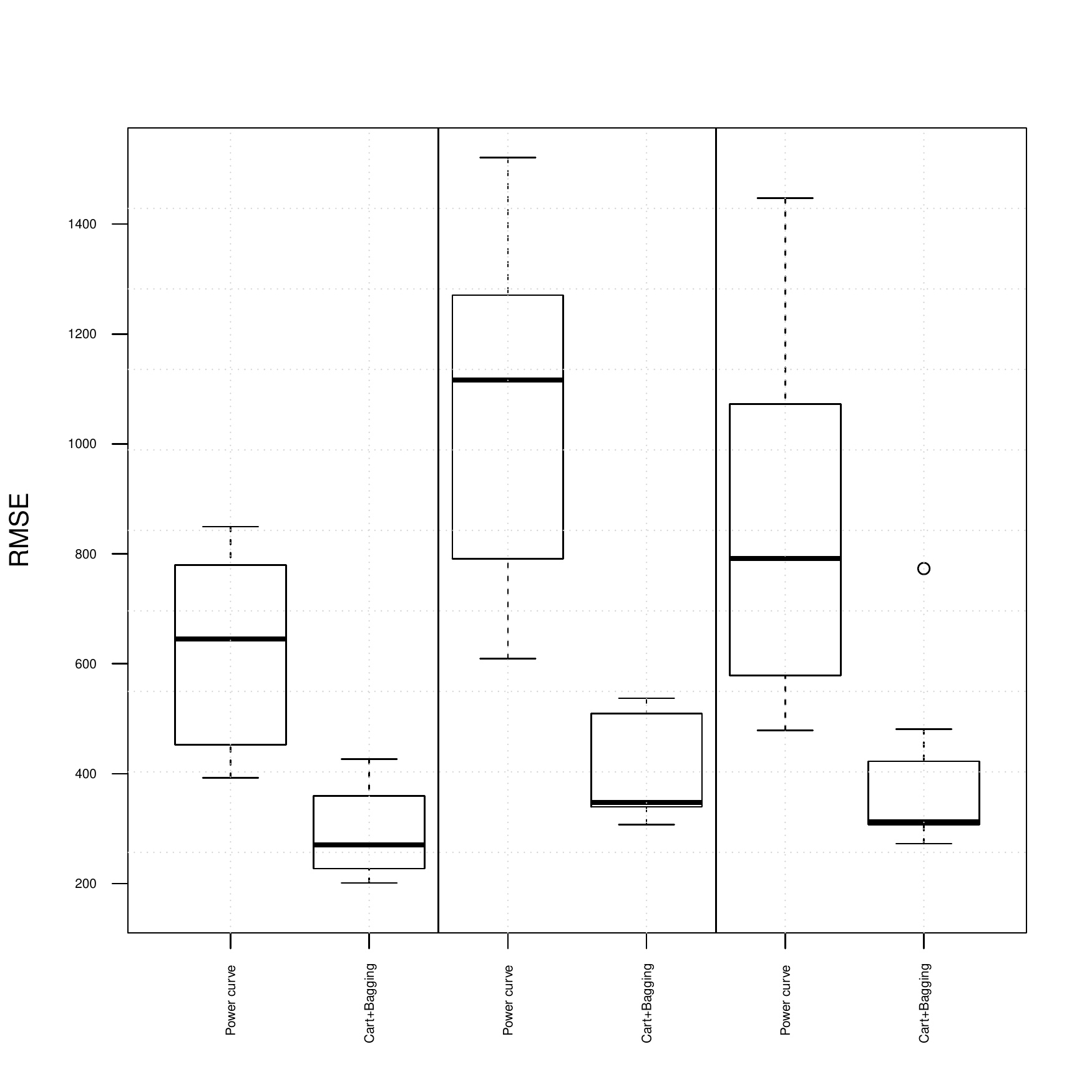}
\caption{Comparison of the RMSE for the turbine builder's power curve and the CART-Bagging procedure on several farms using local measures.}
\label{fig:ComparaisonParcsVmed}
\end{figure}

Just as in the previous framework, CART-Bagging and the turbine builder's power curve prediction have been tested on several farms. Figure~\ref{fig:ComparaisonParcsVmed} stresses the good results of CART-Bagging, which seems robust to the difference between the mean wind speed and the local wind speed on each turbine, contrary to the use of the power curve, suffering from the aggregation of sensor data.

\section{Conclusion and perspectives}

A first interesting conclusion that may be drawn from our study is that, depending on the wind farm, a method which is the best with true local wind information inputs may not perform well any more when using averaged data designed to mimic meteorological wind forecasts. So, despite the good performance of the constructor power curve for one farm, CART-bagging shows to be more robust when turning to aggregated data.

 More generally, this observation raised the following question: in the frame of this work, the data, provided by the company Maïa Eolis, comes from 3 wind farms, all located in the North and East of France, the turbines installed by the company so far being essentially located in these regions, but it could be of prime interest to  have access to wind data from farms in other regions of France, in order to check if there is some universality in the good behavior of the CART-bagging procedure. In other words, does it remain everywhere the best method for aggregated inputs? Note that, in the comparison of data-mining approaches conducted by  
\cite{kusiak2009wind} for US wind farm data, the kNN method, which does not perform particularly well in our study, appears to outperform the other methods, which reinforces the usefulness of continuing our modeling task for other French wind farms.

Here, we calibrated our models using stationary data, that is, data corresponding to full functioning of the wind turbines. 
A complementary work may be to enrich our models including the time slots where wind turbines are working in a non stationary regime (corresponding essentially to start-up regime). This would allow to compute predictions over a (complete) long time of use,  taking into account transitory phenomena of a turbine.

Another direction for future research regards the final step of effective forecasting. In fact, if we get access to meteorological wind forecasts provided by Météo France or the ECMWF as mentioned above, an intermediate step has to be accomplished before simply plugging this information in our models. Indeed, these previsions suffer inevitably from a noticeable bias due to several causes, which has to be corrected in order to build accurate forecasting at the end. For example, the wind speed prevision is provided by the meteorologists for a given height, but the wind speed at the height of the wind turbine may be different.
This shows that it is necessary to elaborate a so-called downscaling method, in other words to find the best possible relationship between real wind at a wind turbine and the meteorological wind forecasts at hand.

%\acks
\section*{Acknowledgements}
We are extremely grateful to Nicolas Girard and Sophie Guignard from the Maïa Eolis Company, for providing the data and for helpful discussions.\\
This research has been supported by a public grant overseen by the French National Research Agency (ANR) as part of the project FOREWER (reference:  ANR-14-CE05-0028).

\bibliographystyle{amsplain}
\bibliography{WindProdForecast} 
\end{document}